\documentclass[11pt]{article}
\usepackage[a4paper,top=3cm,bottom=2cm,left=3cm,right=3cm,marginparwidth=1.75cm]{geometry}
\usepackage{amsmath,amssymb}
\usepackage{indentfirst}
\usepackage{lmodern}
\usepackage{amsfonts}
\usepackage{mathrsfs}
\usepackage{fancyhdr}
\usepackage{mathrsfs,epsfig,morefloats}
\usepackage{natbib}
\usepackage{amsthm}
\usepackage{color}
\usepackage{lscape}
\usepackage{rotating}
\usepackage{caption2}
\usepackage{subfigure}
\usepackage{bm}
\usepackage{enumerate}
\usepackage{colortbl}
\usepackage[colorlinks,
linkcolor=blue,
anchorcolor=blue,
citecolor=blue]{hyperref}
\usepackage[table]{xcolor}
\usepackage{float}
\usepackage{array}
\newtheorem{example}{Example}

\begin{document}
\title{Quantile regression for compositional covariates}
\author{
~~ Xuejun Ma \thanks{ School of Mathematical Sciences, Soochow University, 215006, Suzhou, China, xuejunma@suda.edu.cn}
~~ Ping Zhang \thanks{The corresponding author, School of Mathematical Sciences, Soochow University, 215006, Suzhou, China, 20184207044@stu.suda.edu.cn}
}

\date{}

\maketitle
\begin{abstract}
Quantile regression is a very important tool to explore the relationship between the response variable and its covariates. Motivated by mean regression with LASSO for compositional covariates proposed by \cite{Lin2014}, we consider quantile regression with no-penalty and penalty function. We develop the computational algorithms based on linear programming. Numerical studies indicate that our methods provides the better alternative than mean regression  under many settings, particularly for heavy-tailed or skewed  distribution of the error term. Finally, we study the fat data using the proposed method.
\end{abstract}

\begin{quote}
\end{quote}

\begin{quote}
\noindent
{\sl MSC2010 subject classifications}: Primary 62J05; secondary 62J07.
\end{quote}

\section{Introduction}
\label{section1}
Compositional data are defined traditionally as constrained data, like proportions or percentages, with a fixed constant sum constraint (such as unit-sum constraint), which can find applications in a wide range of geology, sociology, economics,  biology and so on (\cite{Shanmugam2018}). Sometimes, we are especially interested in relative information, not the absolute values, such as geochemical compositions of rocks. Since the seminal work of \cite{Aitchison1982}, statistical methodologies have been proposed for  compositional data analysis. However, owing to the special nature of compositional data, the usual linear regression model is inappropriate for our purposes. The linear log-contrast model of \cite{Aitchison and Bacon-Shone1984} is a very common method for regression to deal with compositional data. To the best of authors' knowledge, statistical methods discussed mean regression. As we known, mean regression is not robust against outliers. For the testing problem, we address quantile regression in the paper.
%For linear models, \cite{Lin2014}  proposed Lasso method for mean regression.

Quantile regression is robust to outliers and heavy-tailed conditional error distributions. Moveover, it can provide a more complete picture than mean regression when the conditional distribution of the response variable  is asymmetric. Hence, quantile regression has been applied in survival analysis, financial economics, investment analysis and so on (\cite{Koenker and Basseet1978}, \cite{Koenker and Geling2001} and \cite{Yuetal2003}).
Variable selection is an important issue in statistical modeling. In recent years, many different types of penalties have been introduced. \cite{Tibshirani1996} proposed LASSO, which imposed the same penalty on every regression coefficient leading to excessive compression of larger coefficients. To tackle the problem, \cite{Fan and Li2001} developed  SCAD, which has  three properties: unbiasedness, sparsity, continuity. \cite{Zou2006} introduced the adaptive LASSO by using adaptive penalizing weights for different coefficients in the LASSO penalty, which meets above three properties. Besides, there are fused LASSO (\cite{Tibshirani2005}),  elastic net (\cite{Zou2005b}) and   MCP (\cite{Zhang2010}). We also refer to \cite{Wu and Liu2009}.

For compositional data, \cite{Lin2014} studied variable selection in mean regression by LASSO. However, this method is  sensitive to outliers. When the error follows  the  heavy-tailed distribution or asymmetric distribution, it works not well.
In this paper, we develop a novel method to overcome these difficulties via combining  quantile regression with the adaptive LASSO penalty, showing the new algorithms based on linear programme.

The paper is organized as follows. In Section 2, we introduce the proposed methods of quantile regression and its variable selection,   and give  a method for selecting the tuning parameter. In Section 3, we present the computational algorithms. Simulation studies and an empirical example are presented in Section 4. The article concludes in Section 5.

\section{Quantile regression  for compositional data}
\label{section2}
Let $(Y_{i},X_{i})$ be an observation collected from the subject $(i=1,\dots,n)$, where $Y_{i}\in R$ is the response of interest, $X_{i}=(X_{i1},\dots,X_{ip})^\top\in R^{p} $  is the p-dimensional covariates  lying in the $(p-1)$-dimensional positive simplex $S^{p-1} = \left\lbrace (X_{i1}, \dots ,X_{ip})|X_{ij}>0,   \sum_{j=1}^{p}X_{ij}=1\right\rbrace $. $Y=(Y_{1}, \dots, Y_{n})^\top$.
We apply the log-ratio transformation of \cite{Aitchison1982}, and lead to  the linear log-contrast model
\begin{equation}\label{eq1}
Y=Z^{p}\beta_{\backslash p} +\varepsilon,
\end{equation}
where $Z^p=\left\lbrace \log (X_{ij}/X_{ip})\right\rbrace $ is an $n \times(p-1)$ log-ratio matrix with the $p$th component as the reference component, $\beta_{\backslash p}=(\beta_1,\dots,\beta_{p-1})^\top$, $\varepsilon$ is an $n$-vector of independent error term.

In the model (\ref{eq1}), the reference component selection is  not easy. As \cite{Lin2014}, let $\beta_p=-\sum_{j=1}^{p-1}\beta_j$,
the expression can be rewritten as
\begin{equation}\label{eq2}
Y=Z\beta +\varepsilon,\quad \sum_{j=1}^{p}\beta_{j}=0,
\end{equation}
where $Z=\{\log x_{ij}\}_{n\times p}=(Z_1, \dots,Z_n)^\top$, $\beta=(\beta_{1},...,\beta_{p})^ \top$. Note that the intercept is not included in the model, since it can be eliminated by centring the response and predictor variables. So, the estimatorof quantile regression  is to minimize the following objective function
\begin{align}\label{eq3}
&\text{arg}\min_{\beta} \sum_{i=1}^{n}\rho_{\tau}(Y_{i}-Z_{i}^\top\beta)\textcolor{red}{,} \\
& s.t. \sum_{j=1}^{p}\beta_{j}=0, \notag
\end{align}
where $\rho_{\tau}(u)=u(\tau-I(u<0))$ is called the check function. $\tau\in(0,1)$ is the quantile, and
$I(\cdot)$ is an indicative function. $\beta=(\beta_{1},\dots,\beta_{p})^\top$ is the unknown parameter vector.

%\section{Quantile regression  with adaptive Lasso }
%\label{section3}
Now, we consider quantile regression  with the adaptive LASSO, which is  motivated by the LASSO penalty mean regression  proposed by \cite{Lin2014}. As we known, the adaptive LASSO penalty function is a generalization of the LASSO penalty via  adaptive weights. Hence, we consider the constrained  optimization problem
\begin{align}\label{eq4}
&\text{arg}\min_{\beta}  \Big{(}\sum_{i=1}^{n}\rho_{\tau}(Y_{i}-Z_{i}^\top\beta) + \lambda \sum_{j=1}^{p}w_{j}\vert\beta_{j} \vert\Big{)}\textcolor{red}{,}\\
& s.t. \sum_{j=1}^{p}\beta_{j}=0, \notag
\end{align}
where $w_{j}=\dfrac{1}{\vert\widetilde{\beta}_{j}\vert^\kappa}$, and $\kappa >0$. ${\widetilde{\beta}}$ is the solution of model (\ref{eq3}). Here, $\lambda$ is the  tuning parameter. As the suggestion of \cite{Zou2006}, we set $\kappa=1$ in our paper.

In the problem (\ref{eq4}), the tuning parameter is very important because the penalty method depends  on the choice of it. We can use the BIC criterion to select the parameter $\lambda$ (\cite{Wang2007}), which is
defined as

$$
 {\rm BIC}(\lambda_n)=\log\Big{(}\sum_{i=1}^{n}\rho_{\tau}(Y_i-Z_i^{\top}\beta)\Big{)}+\dfrac{\log n}{n}\times df
 $$
where $df$ is the number of nonzero coefficients. The optimal regularization parameter $\lambda_{opt} = {\rm arg} \min_{\lambda_n}\rm{BIC}(\lambda_n)$.
\section{Algorithms}
From (\ref{eq3}) and (\ref{eq4}), they are the constrained optimization problems. First of all, we deal with quantile regression with the adaptive LASSO. Here, we give the computational algorithm, which introduces  some slack variables replacing the objective function with an equality constraint so that (\ref{eq4}) can be transformed  into a linear programming problem.

 Let $u_{i}=\max(0,Y_{i}-Z_{i}^\top \beta)$, $v_{i}=\max\big(0,-(Y_{i}-Z_{i}^\top \beta)\big)$, $\beta^{+}_{j}=\max(0, \beta_j)$,   $\beta^{-}_{j}=\max(0, -\beta_j)$.  $\beta_{j}=\beta^{+}_{j}-\beta_{j}^{-}$ and $\vert\beta_{j}\vert=\beta^{+}_{j}+\beta_{j}^{-}$. $w=(w_1,\dots,w_p)^\top$. By the expressions of slack variables, (\ref{eq4}) can be re-expressed as
\begin{align*}
&\sum_{i=1}^{n}\rho_{\tau}(Y_{i}-Z_{i}^\top \beta)  + \lambda \sum_{j=1}^{p}w_{j}\vert \beta_{j}\vert\\
 = &  \sum_{i=1}^{n}\Big{(} \tau u_{i}+(1-\tau) v_{i} \Big{)} + \lambda \sum_{j=1}^{p}w_{j}(\beta_{j}^{+} + \beta_{j}^{-}) \\
 = &  \tau I_{n}^\top u +(1-\tau) I _{n}^\top v + \lambda w^\top \beta^{+} + \lambda w^\top \beta^{-} \\
= &  (\lambda w^\top, \lambda w^\top ,\tau I_{n}^\top ,(1-\tau)I_{n}^\top )((\beta^{+})^{\top}, (\beta^{-})^\top, u^\top, v^\top)^\top\\
    \doteq &  A \gamma
\end{align*}
Here $u=(u_1,\dots,u_n)^\top $ and $v=(v_1,\dots,v_n)^\top$. $\beta^{+}=(\beta_{1}^{+}, \dots, \beta_{p}^{+})^\top $  and $\beta^{-}=(\beta_{1}^{-}, \dots, \beta_{p}^{-})^\top$. $I_{n}$ denotes the $n$-vector of ones. $A=(\lambda w^\top,\lambda w^\top ,\tau I_{n}^\top, (1-\tau)I_{n}^\top )$ and $\gamma=\big((\beta^{+})^{\top}, (\beta^{-})^\top, u^\top, v^\top\big)^\top$.

 The constrained condition is
\begin{equation*}
Y- Z\beta=u-v
\end{equation*}

Elementary calculations show that
$$
\left[
\begin{array}{cccc}
Z &  -Z & E_{n} & -E_{n} \\
\end{array}
\right]
\left[
\begin{array}{c}
\beta^{+} \\
\beta^{-}\\
u \\
v \\
\end{array}
\right]
=Y
$$
where $E_n$ denotes the $n\times n$ identity matrix.

 Let
$$ B=
\begin{bmatrix}
I_{p}^{\top} & - I_{p}^{\top} & 0_{n}^{\top} & -0_{n}^{\top} \\
Z &  - Z & E_{n} & -E_{n} \\
\end{bmatrix}
$$
and
$$
H=(0, Y^\top )^\top.
$$
where $0_{n}$ denotes the $n$-vector of zeros. Combined with $\sum_{j=1}^{p} \beta_{j}=0$, the constrained optimization problem (\ref{eq4}) can be transformed into a linear programming problem
\begin{align*}
\min A \gamma&\\
s.t.& B \gamma=H
\end{align*}

Similarly, quantile regression without penalty model (\ref {eq3}) can also transform into a linear programming problem
\begin{align*}
\min A_1 \gamma_1&\\
s.t.& B_1 \gamma_1=H_1
\end{align*}
where $A_1 = (0_{n}^\top, 0_{n}^\top ,\tau I_{n}^\top ,\ (1-\tau)I_{n}^\top ),\ B_1=B,\ \gamma_1=\gamma$, and $H_1=H$.

\section{Numerical studies}
\subsection{Simulations}
As \cite{Lin2014},  we generate the covariate data in the following way. We  first generate an $n\times p$ data matrix $O=(o_{ij})$ from a multivariate normal distribution $N_{p}(\mu,\Sigma)$, and then obtain the covariate matrix
$X=(x_{ij})$, where $x_{ij}=\exp(o_{ij})/\sum_{k=1}^{p}\exp(o_{ik})$. Here $\mu=(\mu_{1},\dots, \mu_{p})^{\top}$. We repeat 500 times for each setting. The error term $\varepsilon$ is generated from five distributions.\\
\textbf{Case 1}. $\varepsilon \sim N(0,1)$.\\
\textbf{Case 2}. $\varepsilon \sim t(3)$, which is symmetric and heavy-tail distribution.\\
\textbf{Case 3}. $\varepsilon \sim pareto(2,1)$, which is the heavy-tail distribution. \\
\textbf{Case 4}. $\varepsilon \sim gpd(0.2,0,1.2)$, which is the skewed distribution.\\
\textbf{Case 5}. $\varepsilon \sim gev(0.2,3,1.5)$, which is the  extreme value distribution, and the skewed distribution.

In Example \ref{example1}, we examine the performance of mean regression (MR) and quantile regression (QR, $\tau=0.5$).  In Example \ref{example2}, we conduct the Monte Carlo comparisons  for variable selection.
\begin{example}\label{example1}
Let
\begin{equation*}
\mu_{j}=\left\{
\begin{aligned}
&\log(0.5*p),  &j=1,2,3\\
&0,  &{\rm others}\\
\end{aligned}
\right.
\end{equation*}
and $\Sigma=\rho^{\vert i-j\vert}$ with $\rho =0.2$. We set $n$ = \{50, 100, 200, 500\}, and generate the responses according to model (\ref{eq2}) with $\beta=(1, -0.8, 0.6, -1.5, -0.5, 1.2)^\top$.  We evaluate the performance through
the following two criteria:
\begin{enumerate}[(1)]
  \item $b_j=\frac{1}{500}\sum_{m=1}^{500}\vert \hat{\beta}_j^{(m)}-\beta_j\vert$
  \item $\ L_1= \frac{1}{500}\sum_{m=1}^{500} \sum_{j=1}^{p} \vert \hat{\beta}_j^{(m)}-\beta_j\vert$
\end{enumerate}
 Here $\hat{\beta}_{j}^{(m)}$ is the estimator of $\beta_{j}$ based on the $m$-th sample. We compare the performance of quantile  regression (QR) with  mean regression (MR).
\end{example}

 Table \ref{table1}  summarizes the simulation results. We can draw the following conclusions:
 \begin{enumerate}[(1)]
  \item When the error distribution follows the  normal distribution, MR is slightly better than QR. As the sample size increases, the differences between them  are decreasing.
  \item When the error distribution is the heavy-tailed or skewed, QR performs better than MR since these distributions have outliers.
\end{enumerate}

\begin{table}[H]\centering
\caption{Simulation results for Example 1}
\label{table1}
\resizebox{\textwidth}{!}{
\begin{tabular}{@{\extracolsep{5pt}} cccccccccc}
\\[-1.8ex]
\hline \\[-1.8ex]
Distribution & $n$ & Method & $b_1$ & $b_2$ & $b_3$ & $b_4$ & $b_5$ & $b_6$ & $L_1$ \\ \hline \\[-1.8ex]
$N(0,1)$ & 50 & MR & $0.114$ & $0.118$ & $0.118$ & $0.118$ & $0.119$ & $0.114$ & $0.701$ \\
\ & \ & QR & $0.143$ & $0.149$ & $0.142$ & $0.148$ & $0.145$ & $0.143$ & $0.870$  \\
\ & 100 & MR &  $0.074$ & $0.079$ & $0.083$ & $0.082$ & $0.082$ & $0.077$ & $0.477$  \\
\ & \ & QR & $0.095$ & $0.101$ & $0.105$ & $0.105$ & $0.099$ & $0.098$ & $0.603$ \\
\ & 200 & MR & $0.054$ & $0.057$ & $0.058$ & $0.057$ & $0.056$ & $0.054$ & $0.336$   \\
\ & \ & QR & $0.070$ & $0.071$ & $0.074$ & $0.072$ & $0.069$ & $0.068$ & $0.424$ \\
\ & 500 & MR & $0.033$ & $0.035$ & $0.034$ & $0.036$ & $0.037$ & $0.034$ & $0.209$   \\
\ & \ & QR & $0.040$ & $0.044$ & $0.043$ & $0.044$ & $0.045$ & $0.042$ & $0.259$  \\
$t(3)$ & 50 & MR & $0.186$ & $0.191$ & $0.196$ & $0.200$ & $0.202$ & $0.184$ & $1.159$  \\
\ & \ & QR & $0.157$ & $0.167$ & $0.172$ & $0.173$ & $0.166$ & $0.160$ & $0.995$  \\
\ & 100 & MR & $0.128$ & $0.133$ & $0.133$ & $0.139$ & $0.143$ & $0.128$ & $0.803$  \\
\ & \ & QR & $0.110$ & $0.114$ & $0.112$ & $0.113$ & $0.115$ & $0.110$ & $0.674$ \\
\ & 200 & MR & $0.090$ & $0.090$ & $0.092$ & $0.095$ & $0.097$ & $0.088$ & $0.553$  \\
\ & \ & QR & $0.075$ & $0.076$ & $0.077$ & $0.077$ & $0.078$ & $0.074$ & $0.457$  \\
\ & 500 & MR & $0.056$ & $0.058$ & $0.060$ & $0.058$ & $0.062$ & $0.060$ & $0.355$  \\
\ & \ & QR & $0.047$ & $0.048$ & $0.049$ & $0.047$ & $0.049$ & $0.047$ & $0.287$ \\
$pareto(2,1)$ & 50 & MR & $0.198$ & $0.212$ & $0.217$ & $0.221$ & $0.219$ & $0.193$ & $1.261$ \\
\ & \ & QR & $0.077$ & $0.083$ & $0.084$ & $0.080$ & $0.084$ & $0.078$ & $0.486$ \\
\ & 100 & MR & $0.162$ & $0.169$ & $0.173$ & $0.166$ & $0.168$ & $0.160$ & $0.999$ \\
\ & \ & QR & $0.050$ & $0.053$ & $0.053$ & $0.054$ & $0.053$ & $0.053$ & $0.317$ \\
\ & 200 & MR & $0.128$ & $0.130$ & $0.129$ & $0.130$ & $0.134$ & $0.128$ & $0.779$ \\
\ & \ & QR & $0.037$ & $0.039$ & $0.040$ & $0.039$ & $0.039$ & $0.036$ & $0.231$ \\
\ & 500 & MR & $0.079$ & $0.091$ & $0.084$ & $0.081$ & $0.086$ & $0.083$ & $0.506$ \\
\ & \ & QR & $0.024$ & $0.025$ & $0.025$ & $0.025$ & $0.025$ & $0.024$ & $0.147$ \\
$gpd(0.2,0,1.2)$ & 50 & MR & $0.209$ & $0.217$ & $0.224$ & $0.217$ & $0.210$ & $0.206$ & $1.284$ \\
\ & \ & QR & $0.147$ & $0.154$ & $0.155$ & $0.162$ & $0.156$ & $0.149$ & $0.924$  \\
\ & 100 & MR & $0.141$ & $0.150$ & $0.146$ & $0.148$ & $0.143$ & $0.140$ & $0.868$  \\
\ & \ & QR & $0.099$ & $0.109$ & $0.107$ & $0.109$ & $0.104$ & $0.104$ & $0.632$ \\
\ & 200 & MR & $0.104$ & $0.105$ & $0.111$ & $0.107$ & $0.113$ & $0.102$ & $0.641$  \\
\ & \ & QR & $0.071$ & $0.075$ & $0.076$ & $0.075$ & $0.078$ & $0.073$ & $0.448$  \\
\ & 500 & MR & $0.064$ & $0.070$ & $0.068$ & $0.067$ & $0.068$ & $0.068$ & $0.404$  \\
\ & \ & QR & $0.045$ & $0.049$ & $0.048$ & $0.048$ & $0.049$ & $0.045$ & $0.285$  \\
$gev(0.2,3,1.5)$ & 50 & MR & $0.291$ & $0.316$ & $0.323$ & $0.318$ & $0.314$ & $0.312$ & $1.874$  \\
\ & \ & QR & $0.255$ & $0.272$ & $0.279$ & $0.286$ & $0.271$ & $0.264$ & $1.627$  \\
\ & 100 & MR & $0.200$ & $0.213$ & $0.216$ & $0.210$ & $0.217$ & $0.213$ & $1.270$ \\
\ & \ & QR & $0.166$ & $0.188$ & $0.191$ & $0.191$ & $0.186$ & $0.186$ & $1.109$  \\
\ & 200 & MR & $0.146$ & $0.150$ & $0.152$ & $0.157$ & $0.158$ & $0.137$ & $0.900$  \\
\ & \ & QR & $0.128$ & $0.133$ & $0.134$ & $0.128$ & $0.134$ & $0.125$ & $0.781$ \\
\ & 500 & MR & $0.093$ & $0.097$ & $0.096$ & $0.097$ & $0.097$ & $0.093$ & $0.573$ \\
\ & \ & QR & $0.082$ & $0.085$ & $0.081$ & $0.085$ & $0.084$ & $0.079$ & $0.495$ \\
\hline \\[-1.8ex]
\end{tabular}}
\end{table}
\begin{example}\label{example2}
 Let
\begin{equation*}
\mu_{j}=\left\{
\begin{aligned}
&\log(0.5*p),  &j=1,\dots, 5\\
&0,  &{\rm others}\\
\end{aligned}
\right.
\end{equation*}
 and $\beta=(1, -0.8, 0.6, 0, 0, -1.5, -0.5, 1.2, 0,\dots, 0)^\top$. We set $(n, \,p)$=\{(50, 10), (100, 10), (100, 20), (200, 20)\}. To summarize the variable selection results and evaluate estimation accuracy, we consider the following criteria:
 \begin{enumerate}[(1)]
  \item TP: the average number of true positives, which denotes the average number  of the true zero  correctly set to zero.
  \item TN: the average number of true negatives, which denotes the average number  of the true nonzero correctly set to nonzero.
  \item FP: the average number of false positives, which denotes the average number of the true zero incorrectly set to nonzero.
  \item FN: the average number of false negetives, which denotes the average number of the true nonzero incorrectly set to zero.
\end{enumerate}
 We compare three method: quantile regression with  adaptive LASSO (QR-ALA, $\tau=0.5$),  mean regression with  LASSO (MR-LA) and  adaptive LASSO (MR-ALA). The other settings are the same as Example \ref{example1}.
\end{example}

From Tables \ref{table2} and \ref{table3}, we can get the following comments:
 \begin{enumerate}[(1)]
  \item From $L_1$,  QR-ALA is  better than MR-LA and MR-ALA, especially for the heavy-tail or skewed distribution. Even if  $N(0,1)$, QR-ALA is still slightly better,  which implies that QR-ALA is  more accurate. The performances of three methods increase gradually with $n$.
  \item  For variable selection, QR-ALA and MR-ALA  outperform than MR-LA, which is more inclined to set zero coefficients to nonzero since FP is very large.  When the error term follows the normal distribution, MR-ALA is better than QR-ALA. However, when the error term follows  other distributions, QR-ALA  is superior to MR-ALA, which clearly indicates that the proposed method is more efficient.
\end{enumerate}

\begin{table}[H]\centering
\caption{Simulation results for Example 2 (Cases 1--2)}
\label{table2}
\resizebox{\textwidth}{!}{
\begin{tabular}{@{\extracolsep{5pt}} cccccccc}\\
\hline \\[-1.8ex]
Distribution & $(n,\, p)$ & Method & $L_1$ & TP & TN & FP & FN \\\hline \\[-1.8ex]
$N(0,1)$ & (50, 10)& MR-LA &  $1.197$ & $5.952$ & $2.318$ & $1.682$ & $0.048$ \\
\ & \ & MR-ALA & $1.245$ & $5.770$ & $3.856$ & $0.144$ & $0.230$ \\
\ & \ & QR-ALA & $1.160$ & $5.740$ & $3.598$ & $0.402$ & $0.260$ \\
\ & (100, 10)& MR-LA  & $0.915$ & $5.996$ & $2.596$ & $1.404$ & $0.004$ \\
\ & \ & MR-ALA & $1.047$ & $5.952$ & $3.988$ & $0.012$ & $0.048$ \\
\ & \ & QR-ALA & $0.728$ & $5.968$ & $3.752$ & $0.248$ & $0.032$ \\
\ & (100, 20) & MR-LA  & $1.251$ & $5.998$ & $11.154$ & $2.846$ & $0.002$ \\
\ & \ & MR-ALA & $1.048$ & $5.936$ & $13.930$ & $0.070$ & $0.064$ \\
\ & \ & QR-ALA & $0.832$ & $5.950$ & $13.488$ & $0.512$ & $0.050$ \\
\ & (200, 20) & MR-LA  & $0.850$ & $6.000$ & $11.478$ & $2.522$ & $0.000$ \\
\ & \ & MR-ALA & $0.943$ & $5.990$ & $13.998$ & $0.002$ & $0.010$ \\
\ & \ & QR-ALA & $0.511$ & $6.000$ & $13.724$ & $0.276$ & $0.000$ \\	
$t(3)$ & (50, 10)& MR-LA  & $2.139$ & $5.242$ & $2.558$ & $1.442$ & $0.758$ \\
\ & \ & MR-ALA & $1.934$ & $4.920$ & $3.626$ & $0.374$ & $1.080$ \\
\ & \ & QR-ALA & $1.648$ & $5.110$ & $3.702$ & $0.298$ & $0.890$ \\
\ & (100, 10)& MR-LA  & $1.502$ & $5.750$ & $2.746$ & $1.254$ & $0.250$ \\
\ & \ & MR-ALA & $1.426$ & $5.472$ & $3.838$ & $0.162$ & $0.528$ \\
\ & \ & QR-ALA & $0.840$ & $5.878$ & $3.832$ & $0.168$ & $0.122$ \\
\ & (100, 20) & MR-LA   & $2.125$ & $5.418$ & $11.964$ & $2.036$ & $0.582$ \\
\ & \ & MR-ALA & $1.525$ & $5.330$ & $13.566$ & $0.434$ & $0.670$ \\
\ & \ & QR-ALA & $1.029$ & $5.812$ & $13.518$ & $0.482$ & $0.188$ \\
\ & (200, 20) & MR-LA  & $1.492$ & $5.910$ & $12.034$ & $1.966$ & $0.090$ \\
\ & \ & MR-ALA & $1.098$ & $5.848$ & $13.816$ & $0.184$ & $0.152$ \\
\ & \ & QR-ALA & $0.586$ & $5.988$ & $13.818$ & $0.182$ & $0.012$\\	
\hline \\[-1.8ex]
\end{tabular}}
\end{table}

\begin{table}[H]\centering
\caption{Simulation results for Example 2 (Cases 3--5)}
\label{table3}
\resizebox{\textwidth}{!}{
\begin{tabular}{@{\extracolsep{5pt}} cccccccc}
\\[-1.8ex]
\hline \\[-1.8ex]
Distribution & $(n,\, p)$ & Method & $L_1$ & TP & TN & FP & FN \\\hline \\[-1.8ex]
$pareto(2,1)$ & (50, 10)& MR-LA & $2.162$ & $5.132$ & $2.394$ & $1.606$ & $0.868$  \\
\ & \ & MR-ALA & $2.193$ & $4.992$ & $3.538$ & $0.462$ & $1.008$\\
\ & \ & QR-ALA & $0.640$  & $5.898$ & $3.824$ & $0.176$ & $0.102$ \\
\ & (100, 10)& MR-LA & $1.643$ & $5.472$ & $2.514$ & $1.486$ & $0.528$  \\
\ & \ & MR-ALA & $1.750$ & $5.344$ & $3.702$ & $0.298$ & $0.656$ \\
\ & \ & QR-ALA & $0.345$ & $6.000$ & $3.956$ & $0.044$ & $0.000$  \\
\ & (100, 20) & MR-LA & $2.335$ & $5.046$ & $11.156$ & $2.844$ & $0.954$ \\
\ & \ & MR-ALA & $1.959$ & $5.048$ & $13.278$ & $0.722$ & $0.952$ \\
\ & \ & QR-ALA & $0.415$ & $5.998$ & $13.844$ & $0.156$ & $0.002$ \\
\ & (200, 20) & MR-LA & $1.921$ & $5.382$ & $10.794$ & $3.206$ & $0.618$ \\
\ & \ & MR-ALA & $1.582$ & $5.438$ & $13.548$ & $0.452$ & $0.562$ \\
\ & \ & QR-ALA & $0.270$ & $6.000$ & $13.940$ & $0.060$ & $0.000$ \\
$gpd(0.2,0,1.2)$ & (50, 10) & MR-LA & $2.235$ & $5.210$ & $2.382$ & $1.618$ & $0.790$ \\
\ & \ & MR-ALA & $2.241$ & $4.772$ & $3.560$ & $0.440$ & $1.228$ \\
\ & \ & QR-ALA & $1.470$ & $5.356$ & $3.732$ & $0.268$ & $0.644$ \\
\ & (100, 10) & MR-LA & $1.621$ & $5.702$ & $2.312$ & $1.688$ & $0.298$ \\
\ & \ & MR-ALA & $1.665$ & $5.242$ & $3.778$ & $0.222$ & $0.758$ \\
\ & \ & QR-ALA & $0.806$ & $5.896$ & $3.844$ & $0.156$ & $0.104$ \\
\ & (100, 20) & MR-LA & $2.500$ & $4.972$ & $10.716$ & $3.284$ & $1.028$ \\
\ & \ & MR-ALA & $1.894$ & $5.074$ & $13.314$ & $0.686$ & $0.926$ \\
\ & \ & QR-ALA & $0.967$ & $5.860$ & $13.554$ & $0.446$ & $0.140$ \\
\ & (200, 20) & MR-LA & $1.851$ & $5.608$ & $9.502$ & $4.498$ & $0.392$ \\
\ & \ & MR-ALA & $1.347$ & $5.654$ & $13.680$ & $0.320$ & $0.346$ \\
\ & \ & QR-ALA & $0.577$ & $6.000$ & $13.748$ & $0.252$ & $0.000$ \\
$gev(0.2,3,1.5)$ & (50, 10) & MR-LA & $3.186$ & $4.350$ & $2.532$ & $1.468$ & $1.650$ \\
\ & \ & MR-ALA & $3.355$ & $3.956$ & $3.162$ & $0.838$ & $2.044$ \\
\ & \ & QR-ALA & $2.779$ & $3.940$ & $3.696$ & $0.304$ & $2.060$ \\
\ & (100, 10) & MR-LA & $2.409$ & $5.008$ & $2.510$ & $1.490$ & $0.992$ \\
\ & \ & MR-ALA & $2.560$ & $4.234$ & $3.620$ & $0.380$ & $1.766$ \\
\ & \ & QR-ALA & $1.759$ & $4.996$ & $3.826$ & $0.174$ & $1.004$ \\
\ & (100, 20) & MR-LA & $3.500$ & $4.732$ & $9.034$ & $4.966$ & $1.268$ \\
\ & \ & MR-ALA & $2.919$ & $3.996$ & $12.762$ & $1.238$ & $2.004$ \\
\ & \ & QR-ALA & $2.056$ & $4.946$ & $13.358$ & $0.642$ & $1.054$ \\
\ & (200, 20) & MR-LA & $2.477$ & $5.566$ & $8.206$ & $5.794$ & $0.434$ \\
\ & \ & MR-ALA & $2.031$ & $4.666$ & $13.682$ & $0.318$ & $1.334$ \\
\ & \ & QR-ALA & $1.239$ & $5.732$ & $13.468$ & $0.532$ & $0.268$ \\
\hline \\[-1.8ex]
\end{tabular}}
\end{table}

\subsection{Application}

In this section, to illustrate the usefulness of the proposed procedure, we apply the proposed method in the dataset fat, which  contains many physical measurements of 252 males can be found in R package "UsingR''. Body.fat is the response variable. The following X-variables are used as covariates: neck (circumference), chest (circumference), abdomen (circumference), hip (circumference), thigh (circumference), knee (circumference), ankle (circumference), bicep (circumference), forearm (circumference) and wrist (circumference). We transform covariates into compositional data. As the suggestion of \cite{Shanmugam2018},  $Y=\log\big(\rm{body.fat} / (100-\rm{body.fat})\big)$.  There are 251 observations after removing suspicious observations. Here, the ten-fold cross-validation method is used to select the tuning parameter. To  evaluate the performance of  MR-ALA and QR-ALA, we divide the data set into a test set and a training set, 9 copies as the training set, and 1 copy as the test set at random. We repeat 100 simulations and use NMSE to compare two methods. NMSE is defined by
$$
   {\rm NMSE}=\dfrac{\sum(Y_{i}-\hat{Y}_{i})^2}{\sum(Y_{j}-\overline{Y})^2}.
$$
where $\overline{Y}$ is the mean of the response variable, $\hat{Y}_{i}$ is the predictive value of the test data set using the model obtained from the training set.

As Table \ref{table4},  NMSE of QR-ALA is less than MR-ALA whether it is the raw data or transformed compositional data, which QR-ALA is better than that MR-ALA since there are outliers in the dataset fat. It is surprised that the performances of the two methods with compositional data are better than the corresponding models with original data, which implies this transform may be   necessary and meaningful in application.

\begin{table}[H]\centering
\caption{NMSE for dataset fat}
\label{table4}
\begin{tabular}{@{\extracolsep{5pt}} ccc} \\[-1.8ex] \hline \\[-1.8ex]
& MR-ALA & QR-ALA \\ \hline \\[-1.8ex]
Original data      & 0.426 & 0.376  \\
Compositional data & 0.424 & 0.353  \\
\hline \\[-1.8ex]
\end{tabular}
\end{table}

\section{Discussion}
In this paper, we study  quantile regression with compositional data, and propose penalized quantile regression with the adaptive-LASSO penalty function. Due to linear programming,  the proposed of the algorithm   works not well when dimension $p$ is much larger than sample size $n$. This problem may be  achieved by ADMM \cite{Yu and Lin2017} proposed. We will study
it in our future research.

\end{document}